\documentclass[aps,pra,reprint,longbibliography,nofootinbib]{revtex4-2}

\usepackage{amsmath,amssymb,mathtools,bm}
\usepackage{siunitx}
\usepackage{microtype}
\usepackage{graphicx}
\usepackage{booktabs}
\usepackage[dvipsnames]{xcolor}
\usepackage{hyperref}
\usepackage[nameinlink,capitalise]{cleveref}

\hypersetup{colorlinks=true,linkcolor=MidnightBlue,citecolor=MidnightBlue,urlcolor=MidnightBlue}

\newcommand{\br}{\mathbf r}

\begin{document}

%===============================================================================
% Header + Section I (replacement)
%===============================================================================
\title{Radial Gausslets}

\author{Steven R. White}
\affiliation{Department of Physics and Astronomy, University of California, Irvine, Irvine, CA 92697, USA}
\date{\today}

\begin{abstract}
 Gausslets are one of the few examples of basis sets for electronic structure which allow for two-index/diagonal electron-electron interaction terms. 
 A weakness of gausslets is that, because of their 1D origin, they have been tied to Cartesian coordinates. Here we generalize the gausslet construction for the radial coordinate in three dimensions for atomic basis sets. These radial gausslets make a very compact radial basis with a relatively modest number of functions, with diagonal interaction terms.  We illustrate the accuracy of this construction with Hartree--Fock and exact diagonalization on atomic systems. 
\end{abstract}

\maketitle

\section{Introduction and background}
\label{sec:intro}
Gausslets\cite{White17,WhiteStoudenmire2019MultiSliced,QiuWhite2021Hybrid,White2023} are local orthogonal basis functions for electronic structure with a diverse set of desirable properties.  %: orthonormality, locality, simple integral evaluation, smoothness, variable resolution depending on distance from a nucleus, and most importantly, permitting a diagonal approximation for the two-electron interaction. 
As such, they sit alongside a number of real-space and multiresolution approaches in electronic-structure and AMO theory, including discrete-variable representations and related grid/quadrature methods,\cite{Light85} wavelet and multiwavelet formulations,\cite{Flad2002,Harrison2004,Mohr2014} and B-spline bases for radial and continuum problems.\cite{Bachau2001,Zatsarinny2006} What is unusual about gausslets within this broader landscape is the simultaneous combination of a number of properties which tend to conflict: orthonormality, locality, smoothness, variable resolution, and an accurate diagonal approximation for the two-electron interaction. A limitation of gausslets is that their complicated construction---based on special wavelet transformations of a 1D grid of gaussians---is only known in 1D, and so the 3D basis sets generated so far have been in coordinate-product form: $f(x) g(y) h(z)$. A direct 2D/3D construction with the full gausslet property set would likely reduce basis sizes at fixed accuracy and improve efficiency across many electronic-structure methods.
%Gausslets\cite{White17,WhiteStoudenmire2019MultiSliced,QiuWhite2021Hybrid,White2023} are basis functions for electronic structure with a diverse set of desirable properties: orthonormality, locality, simple integral evaluation, smoothness, variable resolution depending on distance from a nucleus, and most importantly, permitting a diagonal approximation for the two-electron interaction. A limitation of gausslets is that their complicated construction---based on special wavelet transformations of a 1D grid of gaussians---is only known in 1D, and so the 3D basis sets generated so far have been in coordinate-product form:  $f(x) g(y) h(z)$. A direct 2D/3D construction with the full gausslet property set would likely reduce basis sizes at fixed accuracy and improve efficiency across many electronic-structure methods. 

As a step towards more general constructions, here we develop radial gausslets.  Radial grids are very common in electronic structure, but radial basis sets with gausslet properties are not. Basis sets have advantages in maintaining a variational principle for ground states and excellent accuracy versus size. Our radial gausslets are based on ordinary gausslets with a few modifications to treat (a) the $r^2$ radial metric and (b) completeness and the boundary condition at $r=0$.  They are used with a flexible coordinate transformation giving any desired resolution of the basis versus $r$. 

In Sec.~II we briefly review standard gausslets. In Sec.~III, we introduce techniques for treating a 1D system with an edge, such as the half-line $r\ge 0$. In Sec.~IV, we introduce radial gausslets and the coordinate transformations needed for variable resolution. In Sec.~V, we discuss combining radial gausslets with spherical harmonics and present Hartree--Fock and exact diagonalization tests for atoms. Sec.~VI contains discussion and outlook.

\section{Review of standard gausslets}

\label{sec:stdgausslets}

Gausslets were introduced in Ref.~\cite{White17} as one-dimensional (1D) basis functions which
combine grid-like and basis-set-like virtues.  They are orthonormal, smooth, and localized, but
they also behave like \emph{grid points} in an important sense: when integrated against a smooth
function they return (to very high order) the value of that function at the gausslet center.
This ``$\delta$-function'' property is the key that permits a two-index/diagonal approximation for
electron--electron interactions.

In this section we briefly summarize the 1D gausslet construction and the resulting properties
most relevant for the radial generalizations developed later.  Further details and extensive tests
are given in Refs.~\cite{White17,WhiteStoudenmire2019MultiSliced,QiuWhite2021Hybrid,White2023}.

\subsection{Uniform 1D gausslets}

A \emph{uniform} gausslet basis is defined on an infinite 1D lattice with spacing $a$.
One chooses a single normalized ``mother'' gausslet $G(x)$ with unit lattice spacing and then forms
translated and scaled functions
\begin{equation}
  G_i(x) = \frac{1}{\sqrt{a}}\, G\!\left(\frac{x}{a}-i\right),
  \qquad i \in \mathbb{Z}.
  \label{eq:Gi_def}
\end{equation}
The basis is orthonormal,
\begin{equation}
  \int_{-\infty}^{\infty} dx\; G_i(x)\,G_j(x) = \delta_{ij},
  \label{eq:orthonorm}
\end{equation}
and each $G_i(x)$ is strongly localized around $x_i = ia$.

The defining practical feature of gausslets is that they are \emph{explicitly} representable as a
short linear combination of Gaussians on a finer auxiliary grid\cite{White17} with spacing $a/3$. We have
\begin{equation}
  G(x) = \sum_{j} b_j \exp\!\left[-\frac{1}{2}(3x-j)^2\right]
  \label{eq:gausslet_gaussian_sum}
\end{equation}
with $b_{-j}=b_j$ and with the sum effectively truncated because the coefficients are tiny outside a
moderate range of $j$.\footnote{In practice, the coefficients are tabulated once and reused; see
Ref.~\cite{White17}.}  This representation makes overlap, kinetic-energy, and (Gaussian-decomposed)
Coulomb integrals straightforward, because all required integrals reduce to sums of analytic Gaussian
integrals.

The order of a gausslet (e.g.\ $G6$, $G8$, $G10$ in the notation of Ref.~\cite{White17}) controls how
well the basis represents smooth functions; higher order gives higher polynomial completeness at the
price of a modest increase in spatial extent.  We almost always use $G10$, giving 10th order polynomial completeness.

\subsection{Polynomial completeness, COMX, and the moment property}

Gausslets are designed to be excellent at representing smooth functions.  A convenient way to state
this is in terms of \emph{polynomial completeness}: over the spatial region of interest, linear
combinations of $\{G_i\}$ can represent low-degree polynomials essentially exactly (in practice, to
near machine precision for reasonable $a$)~\cite{White17,White2023}.

More special than completeness is a set of \emph{moment} conditions.  Define the ``weight''
\begin{equation}
  w_i \equiv \int dx\; G_i(x),
  \label{eq:wi_def}
\end{equation}
which is a constant $w_i=\sqrt{a}$ for uniform gausslets, but which varies with $i$ for distorted gausslets (below).
Then for low integer $m$ (up to an order set by the gausslet construction), gausslets satisfy
\begin{equation}
  \int dx\; G_i(x)\,(x-x_i)^m = w_i\,\delta_{m0}.
  \label{eq:moments}
\end{equation}
Equivalently, for any polynomial $P(x)$ of sufficiently low degree,
\begin{equation}
  \int dx\; G_i(x)\,P(x) = w_i\,P(x_i).
  \label{eq:delta_property}
\end{equation}
This is the sense in which a gausslet behaves like a $\delta$-function located at $x_i$.

Closely related is the ``X'' property: the coordinate operator is diagonal in the gausslet basis,
\begin{equation}
  X_{ij} \equiv \int dx\; G_i(x)\,x\,G_j(x) = x_i\,\delta_{ij}.
  \label{eq:Xdiag}
\end{equation}
For uniform gausslets this follows very directly from their exact even symmetry about their centers:
$G_i(x)$ is even about $x_i$, and for $i\neq j$ the product $G_i(x)G_j(x)$ is even about the midpoint
$(x_i+x_j)/2$, so the integral of $(x-(x_i+x_j)/2)G_iG_j$ vanishes.  Together with orthogonality, this
gives Eq.~(\ref{eq:Xdiag}).  In the language of Ref.~\cite{White2023}, gausslets are (for practical
purposes) COMX: complete, orthonormal, moment-satisfying, and $x$-diagonalizing.

For general 1D basis sets, the COMX theorem\cite{White2023} is of central importance: if the basis has properties C, O, and X, then it automatically has the moment properties M.  We will make use of the COMX theorem in constructing radial gausslets. 

The practical interpretation is simple: the set $\{(x_i,w_i)\}$ behaves like a high-order quadrature
rule for smooth functions, with the gausslets providing a localized orthonormal ``real-space'' basis
that realizes that quadrature.

\subsection{Diagonal approximations for local operators and for Coulomb interactions}

Consider a one-particle potential $U(x)$.  Its matrix elements are
\begin{equation}
  U_{ij} = \int dx\; G_i(x)\,U(x)\,G_j(x).
  \label{eq:Uij_full}
\end{equation}
Using the moment property, one obtains a diagonal approximation of the form
\begin{equation}
  U_{ij} \;\approx\; \delta_{ij}\,U(x_i),
  \label{eq:Uij_point}
\end{equation}
with closely related ``integral'' and ``summed'' diagonal variants discussed in Ref.~\cite{White17}. In practice, the integral diagonal approximation (IDA) is the most useful and accurate, with 
\begin{equation}
U_{ij} \;\approx\; \frac{\delta_{ij}}{w_i}\int dx\; G_i(x) U(x) .
\label{eq:Uij_ida}
\end{equation}
The key point is not that $U_{ij}$ is numerically small away from the diagonal (although locality
helps), but that the action of $U$ on smooth wavefunctions expanded in gausslets is well captured by the diagonal form.

The same idea extends to the two-electron interaction.  In a generic basis $\{\phi_i\}$, the Coulomb
integrals form a rank-4 tensor
\begin{equation}
  (ij|kl) = \int d\mathbf{r}\, d\mathbf{r}'\;
  \frac{\phi_i(\mathbf{r})\,\phi_j(\mathbf{r})\,
        \phi_k(\mathbf{r}')\,\phi_l(\mathbf{r}')}{|\mathbf{r}-\mathbf{r}'|}.
  \label{eq:eri_full}
\end{equation}
For a gausslet (or gausslet-product) basis, the moment property implies a diagonal approximation
\begin{equation}
  (ij|kl) \;\approx\; \delta_{ij}\,\delta_{kl}\,V_{ik},
  \label{eq:eri_diag}
\end{equation}
where $V_{ik}$ is a two-index interaction matrix. Again, this equation is not meant to indicate that off-diagonal integrals in Eq. (\ref{eq:eri_full}) are very small, but that the replacement of the full $(ij|kl)$ by the diagonal form as a whole is very accurate. In the simplest ``point'' form,
$V_{ik}\approx |\mathbf{r}_i-\mathbf{r}_k|^{-1}$, with $\mathbf{r}_i$ the center of basis function $i$.
Again, the IDA form is the most accurate and useful:
\begin{equation}
 V_{ik} = \frac{1}{w_i w_k} \int d\mathbf{r}\, d\mathbf{r}'\;
  \frac{\phi_i(\mathbf{r})\,
        \phi_k(\mathbf{r}')}{|\mathbf{r}-\mathbf{r}'|}.
  \label{eq:ida_diag}
\end{equation}
The reduction from $\mathcal{O}(N^4)$ to $\mathcal{O}(N^2)$ terms is
immediate, and the resulting second-quantized interaction becomes a sum of density--density terms.
This is particularly advantageous for methods whose cost depends strongly on the number of distinct
two-electron operator terms, such as DMRG and related tensor network methods~\cite{White17,WhiteStoudenmire2019MultiSliced}.

The diagonal approximation is not variational, but in practice the IDA errors are much smaller than errors due to incompleteness of the basis in resolving the electron-electron cusp.

\subsection{Variable resolution and 3D constructions}

Although gausslets originate as uniform 1D objects, one can obtain variable resolution using a
coordinate mapping $t(x)$ with density $\rho(x)\equiv dt/dx$.
A distorted gausslet is defined by the change of variables
\begin{equation}
  \tilde G_i(x) = \sqrt{\rho(x)}\,G\!\left(t(x)-i\right),
  \label{eq:distorted_gausslet}
\end{equation}
which preserves orthonormality. The gausslets are uniform in $t$-space but distorted in $x$. If $\rho(x)$ varies slowly on the scale of the gausslets, the
moment property and the associated diagonal approximations remain accurate~\cite{WhiteStoudenmire2019MultiSliced,White2023}.
This is the basic tool for concentrating basis functions near nuclei while keeping the basis modest
elsewhere.

Existing 3D gausslet bases have been built from these 1D ingredients.  The simplest construction is
a coordinate-product form
\begin{equation}
  \Phi_{ijk}(x,y,z)=\tilde G_i(x)\,\tilde G_j(y)\,\tilde G_k(z),
  \label{eq:product3d}
\end{equation}
which makes many integral manipulations efficient but ties the construction to Cartesian axes.
More flexible Cartesian schemes include multislicing~\cite{WhiteStoudenmire2019MultiSliced}, where the
mapping in one coordinate direction depends on the position in previously-sliced directions, and the
nested gausslet constructions of Ref.~\cite{White2023}, which greatly extend the achievable range of
resolutions while avoiding artifacts of simple product mappings.

The central motivation for this paper is that, despite these advances, the underlying 1D origin of
gausslets has so far kept them closely tied to $x,y,z$ product structures.  For atoms and other
nearly spherical environments, it is natural to ask for a radial analogue that retains the key
gausslet properties---orthonormality, locality, smoothness, variable resolution, and a diagonal
two-electron interaction representation---while treating the radial coordinate and its boundary and
metric factors in a direct way.  We begin addressing those issues in the next section by discussing
gausslet bases on domains with an edge.

\section{Boundary Gausslets}
When we move from an infinite uniform grid to a domain with a \emph{boundary}, all of the nice properties of gausslets need to be revisited. The simplest thought experiment is to take a one-dimensional uniform gausslet basis on the full line, truncate it to $x>0$ by multiplying every function by a step function $\Theta(x)$, and then throw away all centers with $x_i<0$. This does produce a set of localized functions on $[0,\infty)$, but it is not a good basis. Orthogonality is spoiled because cutting off the left-hand tails changes the overlaps for gausslets near the boundary. In addition, the low-order polynomial completeness and moment properties are lost. 

The first step in addressing these problems is fixing the incompleteness.  Without the boundary, for small positive $x$, the completeness comes about from functions in the immediate vicinity of $x$, which includes functions with centers less than zero. To restore completeness, we add those back into the basis, again multiplied by $\Theta(x)$.  The exponential locality of the gausslets means that functions far to the left of the origin are not needed, so we add a small number $N_t$ of extra ``tail'' functions.  As $N_t$ is increased, the overlap matrix of the functions becomes singular, indicating the extra functions are not contributing, and causing inaccuracies requiring higher precision. In our typical unit-space construction we automatically include the gausslet at $x_i=0$, so the extra tail functions stem from 
$x_i=-N_t,-N_t+1,\ldots -1$.  We find a practical limit to be $N_t \le 7$ for double precision calculations.

Once the tails are specified, the COMX procedure and theorem do the rest. First we verify that the COMX theorem applies to a half-line or finite interval. Fortunately, the COMX proof is simple.  Start from a set of functions $\{\Phi_i(x)\}$ defined on an arbitrary finite, semi-infinite, or infinite interval,  that are (i) orthonormal and (ii) polynomially complete up to some order $p$. Build the position matrix
\begin{equation}
X_{ij} = \int \Phi_i(x)\,x\,\Phi_j(x)\,dx
\end{equation}
with all integrals over the specified domain, and diagonalize it to get eigenvalues $x_m$ and eigenvectors $\psi_m$.
The eigenfunctions satisfy
\begin{equation}
\int \psi_m(x)\,x\,\psi_n(x)\,dx = x_m\,\delta_{mn}.\label{Eq:Xeq}
\end{equation}
Using Eq. (\ref{Eq:Xeq}) and orthonormality, we obtain
\begin{equation}
\int \psi_m(x)\,(x-x_m)\,\psi_n(x)\,dx = 0.\label{Eq:Xeq2}
\end{equation}
By completeness,  any polynomial $q(x)$ of degree up to $p$,  can be expanded in the $\psi_n$ basis, so Eq. (\ref{Eq:Xeq2}) implies
\begin{equation}
\int \psi_m(x)\,(x - x_m)\,q(x)\,dx = 0.
\end{equation}
Taking $q(x)=1$ gives the first centered moment condition
\begin{equation}
\int \psi_m(x)\,(x - x_m)\,dx \;\approx\; 0,
\end{equation}
and higher choices of $q$ give higher-order centered moments. This is the sense in which $C+O+X$ \emph{force} $M$: once the span can represent $(x - x_m)q(x)$ for low-degree $q$, the eigenvalue property of $X$ drives the centered moments to zero.

\begin{figure}[t]
\includegraphics[width=0.8\columnwidth]{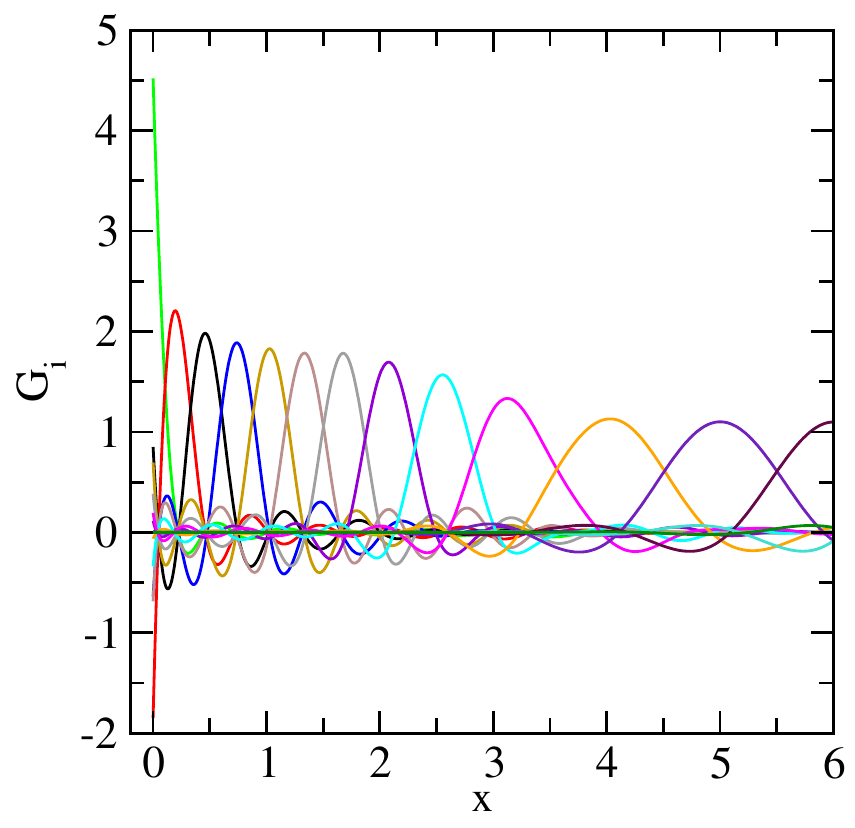}
\caption{Boundary gausslets, where $N_t=6$ extra edge gausslet tails were added to ensure completeness near $x=0$.  The gausslets become noticeably similar to their original form in location and shape for $x \ge 4$.  }
\label{fig:boundarygausslets}
\end{figure}

We take the truncated-and-tailed set on $[0,\infty)$, construct its overlap matrix $S$ and coordinate matrix $X$ with integrals restricted to $x>0$, orthonormalize by $S^{-1/2}$, and then diagonalize $X$ in that orthonormal frame. The resulting $X$-localized modes form what we will call \emph{boundary gausslets}: they are orthonormal on the half-line, sharply localized near their coordinate eigenvalues, and---thanks to the COMX mechanism---retain the $\delta$-moment property to essentially the same order as the original full-line basis. In the interior these boundary gausslets look almost indistinguishable from the original gausslets; near the edge they adjust  to accommodate the additional tail functions. Figure 1 shows boundary gausslets for the half-line.  We find that the boundary gausslets are essentially indistinguishable from the original gausslets if their center $x_i \ge 20$.

A coordinate transformation (if desired) to distort the basis is performed after the boundary gausslet construction. Thus for fixed $N_t$ the set is universal and coefficients in terms of the original gausslets, or the underlying array of gaussians, can be tabulated (although the whole construction is quick).

It is important to note that the general-purpose boundary gausslets of this section need modification before they can be applied for radial functions, as discussed in the next section.

\section{Radial Gausslets}
\subsection{Radial metric and boundary condition}
For central potentials in 3D the radial metric $r^2$ implies that instead of representing the radial function $R(r)$ in terms of gausslets, we should use them for the reduced radial function
\begin{equation}
u(r) = r R(r)
\end{equation}
so that normalization and inner products are expressed in metric-free integrals such as 
\begin{equation}
\int_0^\infty |u(r)|^2\,dr.
\end{equation}
In this form one could immediately introduce boundary gausslets, except that we know that $R(r)$ is finite at the origin, so $u(r=0)=0$.  Rather than imposing an inconvenient constraint on the wavefunction coefficients, we incorporate this boundary condition directly into the basis, so that every basis function $\chi_i(r)$
satisfies
\begin{equation}
\chi_a(0) = 0.
\end{equation}

A natural way to impose vanishing boundary conditions is to start from the non-vanishing boundary basis and remove only the one direction in coefficient space that controls the \emph{value} at the edge. This can be described as fitting $\delta(r)$ with the basis, orthogonally rotating the basis so that the $\delta$ vector is one function, and removing the function, leaving  the rest orthogonal to it. This ``$\delta$-function surgery''
spoils localization and moment properties near the edge, and we must repeat the X diagonalization (and thus could have omitted it initially).   However, it turns out the precise moment property is lost near the edge, due to a failure in the assumptions of the COMX theorem.

After the $\delta$-function surgery, this argument fails at exactly one point: the constant function is no longer in the span, and the proof fails.  Operationally, you can see this by comparing two natural notions of a ``center'' for each boundary gausslet $\psi_m(x)$, the eigenvalue $x_m$, and the moment center
\begin{equation}
\bar x_m = \frac{1}{w_m} \int_0^\infty x\,\psi_m(x)\,dx 
\end{equation}
where $w_m = \int_0^\infty \psi_m(x)\,dx$.
On the full line with exact COMX or standard gausslets these two definitions coincide. After the boundary projection they differ. In practice, these two centers differ significantly only  for the few modes closest to the boundary. Thus the failure of $M$ is tightly localized near the edge, and it shows up precisely as a small mismatch between X-centers and first-moment centers. So we conclude that having \emph{perfect} $M$ seems impossible in this approach without restoring a constant-like degree of freedom.  However, as a practical matter for the radial construction, the differences between the two centers are not large, and they are confined to functions near the origin.  Fig. 2 shows radial gausslets (with no coordinate mapping) near the origin. The largest visible difference between the boundary and radial sets is that the function peaked at the origin is missing. In addition, all the functions vanish at the origin. The functions remain well localized, with relatively small tails, resembling ordinary gausslets with a small coordinate-mapping change of scale.  The figure shows the moments as vertical lines; a slight deviation from equality between the two moments is visible for functions centered at $x < 4$. After a coordinate transformation, this region where there are deviations would be at very small $r$, and its effect not very large, but even that small effect can be minimized significantly.

To quantify the deviations of the centers from each other, we define the merit function
\begin{equation}
D = \sum_i (x_i - \bar x_i)^2.
\label{Eq:Ddef}
\end{equation}
The violation of the COMX theorem stemming from the $r=0$ boundary conditions is tied to $D \ne 0$, and nonzero $D$ stems from deviations of the first, say, ten functions. Even if we had $D=0$, this is only a first order condition, and the functions would still not match the high order zero moments of gausslets on the full line,  but it is a practical scalar measure of the leading first-moment defect, and empirically it tracks the remaining integral diagonal approximation error very well. To make IDA more accurate, we can try to minimize $D$. For the basis shown in Fig. 2, $D \approx 0.011$.

We can make $D$ smaller simply by changing the overall spacing of the basis, or similarly through a coordinate transformation, which one would normally do anyway to allow higher resolution at the core.  The cost is that in order to make the IDA sufficiently accurate, the basis must be made somewhat larger than it would have been if one hadn't had a nonzero $D$. Is a nonzero $D$ unavoidable?  Can modifications of the construction decrease $D$? We have found $D$ can be improved substantially with two modifications of the construction.
\subsection{Odd-Even Construction}
As a first step, we improve the radial construction by exploiting the fact that ordinary gausslets are even about their center.  Let $G_j(x)$ denote uniform unit-spaced gausslets on the full line.  We restrict them to the half line with the step function $\Theta(x)$ and form \emph{odd} and \emph{even} combinations
\begin{subequations}
\begin{align}
O_k(x) &= \Theta(x)\left[G_k(x)-G_{-k}(x)\right], \ \  k=1,2,\ldots, \label{Eq:odddef}\\
E_k(x) &= \Theta(x)\left[G_k(x)+G_{-k}(x)\right], \ \  k=0,1,2,\ldots. \label{Eq:evendef}
\end{align}
\end{subequations}
We focus on the odd combinations first, which automatically satisfy $O_k(0)=0$. They can represent any odd function, so specifically the odd polynomials up to their completeness order. This is the key improvement: odd completeness comes essentially for free.  For large $k$, $O_k(x) \approx G_k(x) \approx E_k(x)$, so the construction only needs to go out to about $k\sim 20$, after which we use $G_k$.  We need to restore the even polynomials for completeness, excluding the constant term, and we do this from the $E_k$.  The significant contributions come from $E_k$ with small $k$, so we place a limit:  we add evens for $k \le K$.  We perform the $\delta$-surgery on this restricted set of $E_k$ to enforce the boundary condition, add the resulting $K-1$ functions to the basis, orthogonalize, and X-localize. Typically we take $K=6$.   We call this improved version the odd-even construction. Note that switching to odd-even still leaves $D \approx 10^{-2}$. 

\begin{figure}[t]
\includegraphics[width=0.8\columnwidth]{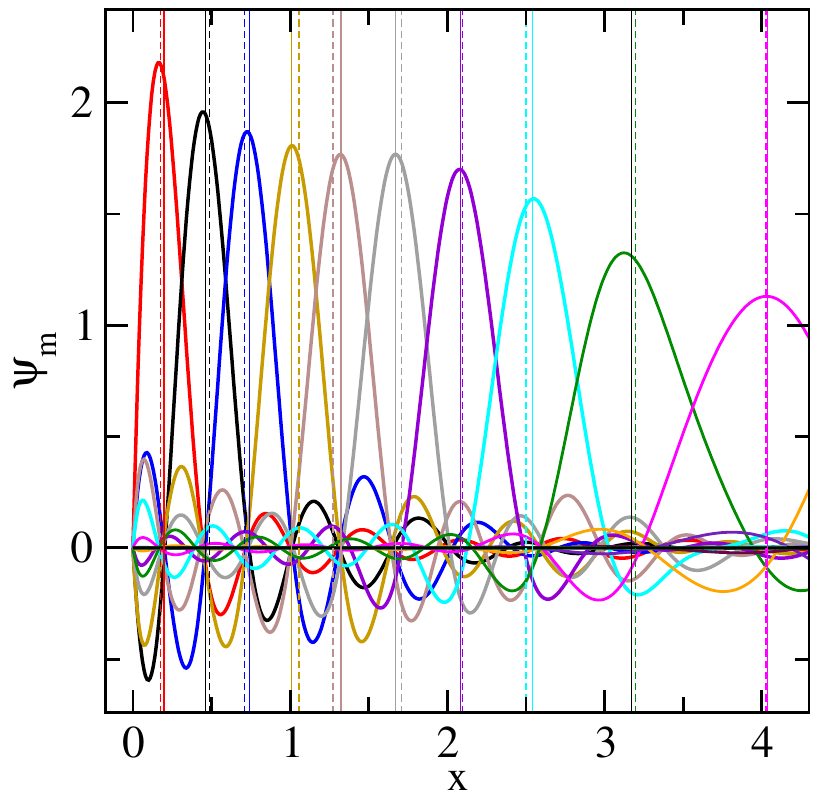}
\caption{Radial gausslets, formed from the boundary gausslets shown in Fig. 1 for $N_t=6$ but with the fit to $\delta(x)$ removed. All the functions vanish at $x=0$. The dashed vertical line for each function shows $\bar x_i$, while the solid line shows $x_i$. These two moments are exactly equal for the boundary gausslets. For $x \ge 4$, the radial gausslets are very similar to the boundary gausslets and $x_i \approx \bar x_i$.  }
\label{fig:radialgausslets}
\end{figure}

\subsection{Adding x-Gaussians}
The second improvement specifically targeting improving $D$ is to add a few additional functions to the odd-even basis, in the form of narrow functions peaked near but slightly greater than $r=0$, and optimize their parameters to minimize $D$. Since the basis is already very complete, the new functions do not need to improve completeness significantly; rather, they provide additional near-origin flexibility to reduce $D$.   We choose functions to add of form 
\begin{equation}
g(x) = x \exp[-\frac{1}{2}(x/\alpha)^2],
\label{Eq:gdef}
\end{equation}
which satisfies the boundary condition, and which we call $x-Gaussians$. We find that adding one or two $x$-Gaussians with optimal $\alpha$'s, chosen by a Nelder--Mead nonlinear optimization, reduces $D$ significantly.  In Fig. 3 we show the resulting basis with $K=6$ and adding the optimal two x-Gaussians.  The agreement with the two centers is very noticeably improved for every function, the functions are still local with small tails, looking like they would form an excellent basis for a diagonal approximation. The resulting $D \approx 1.2\times10^{-5}$, a three orders of magnitude improvement.  The optimal parameters in this case are $\alpha_1=0.0936$ and $\alpha_2=0.0236$, which are roughly comparable to the widths of the added functions. Adding the two x-gaussians allows us to reduce the range of the coordinate transformation, reducing the basis size, while still maintaining good moment properties. \emph{We adopt $K=6$, with these two added x-Gaussians, as our standard radial construction. }

\begin{figure}[t]
\includegraphics[width=0.8\columnwidth]{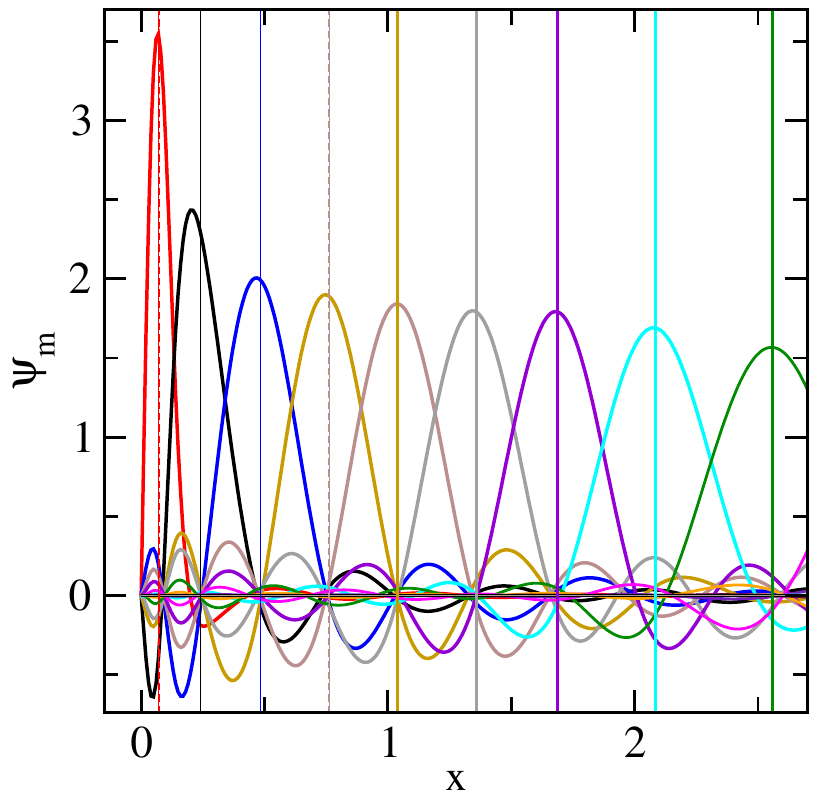}
\caption{Radial gausslets including two extra x-gaussians added; otherwise, similar to Fig. 2. The widths of the two Gaussians were optimized to make all peaks have minimal separation between $x_i$ and  $\bar x_i$ }
\label{fig:radialgaugau}
\end{figure}

\begin{figure}[t]
\includegraphics[width=0.8\columnwidth]{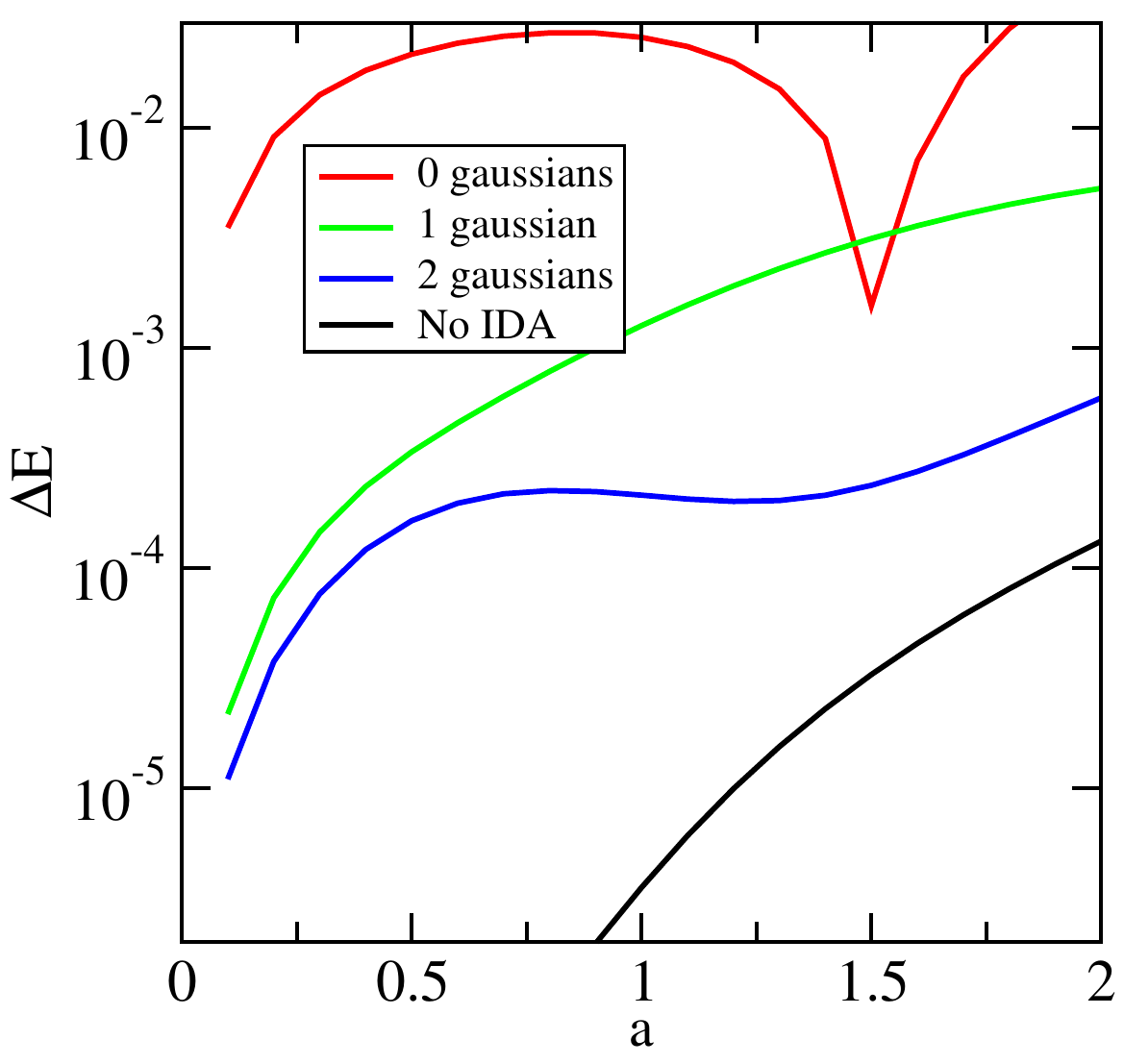}
\caption{Error in the energy of the hydrogen atom versus baseline gausslet spacing for several forms of radial gausslets, when IDA is used for the one--particle potential. All basis functions were scaled uniformly by $a$. The red curve shows the odd-even basis without any extra x-gaussians. The green and blue show with one or two optimized x-gaussians added.  The black curve is with the exact matrix elements of the basis, i.e. the Galerkin form of the potential. }
\label{fig:errorHatom}
\end{figure}

As a test of the diagonal approximation, we apply IDA to the nuclear potential of a hydrogen atom. Normally, IDA is used only for the two-electron interaction, because that is the costly part. Since the nuclear cusp is sharper than the electron-electron cusp, this is a more stringent test than the usual interaction-only use of IDA. In Fig. 4, we see that even without the added gaussians, the radial basis IDA has errors around $10^{-2}$ for unit spacing, $a=1.0$.  It can be made more accurate by making $a$ small, as shown by the drop around $a=0.1$. A much larger improvement is obtained by adding one or two extra Gaussians.  This verifies that minimizing $D$ does significantly improve the IDA. The high accuracy without IDA shows that the basis is very complete, with errors almost entirely due to IDA. It also shows why one should not use IDA everywhere, only for the costly two-electron interaction.

A coordinate mapping as discussed in Section II D is crucial in general since the relevant length scales near the nucleus are much smaller than those in the outer tail of the atom. This means that we naturally tend to be on the far left regime of Fig. 4, with very small IDA errors. In all the tests here we distort the radial coordinate with a specific smooth family of maps, parametrized as
\begin{equation}
t(r) = \frac{1}{s}\,\sinh^{-1}\!\frac{r}{a} + r/10,
\qquad
\rho(r) = \frac{dt}{dr},
\end{equation}
where $a$ changes primarily the core resolution subject to the overall resolution parameter $s$.  The fixed constant 10 is the maximum spacing between gausslets as $r\to \infty$. Near the origin the spacing in $r$ is of order $c=a s$, so in practice we choose $c$ and $s$, and let $a=c/s$. The mapping acts on the radial basis preserving orthonormality: if $\psi_m(t)$ is a radial gausslet on the uniform $t$-axis, we define the physical radial basis function
\begin{equation}
\chi_m(r) = \sqrt{\rho(r)}\,\psi_m\!\big(t(r)\big).
\end{equation}
We expand reduced radial functions $u(r)$ in terms of the $\chi_m(r)$. The derivatives $\chi_m'(r)$ then follow by the chain rule and include both $\psi_m'(t)$ and the Jacobian factor $\rho(r)$. This combination of vanishing-at-the-origin boundary modes and a sinh distortion gives us a compact, orthonormal radial basis $\{\chi_m(r)\}$ that automatically enforces $u(0)=0$ and concentrates resolution near the nucleus.

\section{Hartree--Fock for Atoms}
\subsection{Pure radial case}
In constructing the Hamiltonian, all one-particle radial matrix elements are taken in an exact (Galerkin) form on this basis: no diagonal approximations are used, corresponding to standard practice with conventional gausslet bases. The diagonal approximation, almost always in the IDA form, is used only for the two--electron integrals. For example, suppose we consider a reduced $s$-wave problem where the angular functions are assumed constant.   The single--particle Hamiltonian is
\begin{equation}
H^{(1)} = -\tfrac12 \frac{d^2}{dr^2} - \frac{Z}{r}.
\end{equation}
The overlap matrix is 
\begin{equation}
S_{ab} = \int_0^\infty \chi_a(r)\chi_b(r)\,dr = \delta_{a,b},
\end{equation}
The kinetic matrix is
\begin{equation}
T_{ab} = \frac12\int_0^\infty \chi_a'(r)\chi_b'(r)\,dr,
\end{equation}
and the nuclear attraction is
\begin{equation}
V^{\rm nuc}_{ab} = -Z\int_0^\infty \frac{\chi_a(r)\chi_b(r)}{r}\,dr.
\end{equation}
We evaluate these integrals using a separate coordinate-transformed quadrature grid of $N_g$ points,
with spacing much finer than the radial gausslet spacing but chosen to mirror the same mapping $t(r)$.
The quadrature weights are entirely conventional. This cleanly separates \emph{quadrature accuracy}
from the \emph{diagonal (IDA) approximation}: we can hold the quadrature fixed at high accuracy,
while using a much smaller radial basis which needs only to ensure completeness and an accurate diagonal representation of the two--electron
interaction.  This is an important distinction from discrete variable representations (DVRs), where
the same set of points simultaneously defines the basis and supplies the quadrature, so the ``grid''
must do double duty. The singular nature of the Coulomb interaction, both at small $r$ and along $r=r'$, demands high resolution for accurate quadrature. 

As a first realistic test, we calculate the restricted Hartree-Fock (RHF)  for the helium atom in a  treatment with only the radial degree of freedom (i.e.\ a constant angular function). Here we utilize IDA  for the electron--electron interaction.  
In this spherically symmetric, $S$-only setting, the relevant radial Coulomb kernel arises from integrating the full Coulomb interaction over the angles. If we take a constant angular function (or equivalently the $Y_{00}$ spherical harmonic), then
\begin{equation}
K(r,r') = \frac{1}{(4\pi)^2}
\int d\Omega \int d\Omega'\,\frac{1}{|\mathbf r - \mathbf r'|}
= \frac{1}{\max(r,r')}.
\end{equation}
Given an orthonormal set $\{\chi_a(r)\}$, we define radial weights
\begin{equation}
w_a^\chi = \int_0^\infty \chi_a(r)\,dr
\end{equation}
and construct a two-index interaction matrix using IDA,
\begin{equation}
V_{ab}
= \frac{1}{w_a^\chi w_b^\chi}
\int_0^\infty\!\!\int_0^\infty
\frac{\chi_a(r)\chi_b(r')}{\max(r,r')}\,dr\,dr'.
\end{equation}
This double integral is evaluated on the same coordinate-transformed grid used for the one-particle terms. There is a standard trick that makes the computation much faster than a naive two-dimensional quadrature: we split the domain into regions with $r' < r$ and $r' > r$, introduce prefix integrals
\begin{equation}
I_b(r) = \int_0^r \chi_b(s)\,ds,
\end{equation}
and rewrite the double integral in terms of one-dimensional integrals involving $\chi_a(r)$, $I_b(r)$ and their roles interchanged. The result is an $O(N_{\mathrm{grid}})$ procedure per pair $(a,b)$, rather than an $O(N_{\mathrm{grid}}^2)$ one.

\begin{figure}[t]
\includegraphics[width=0.8\columnwidth]{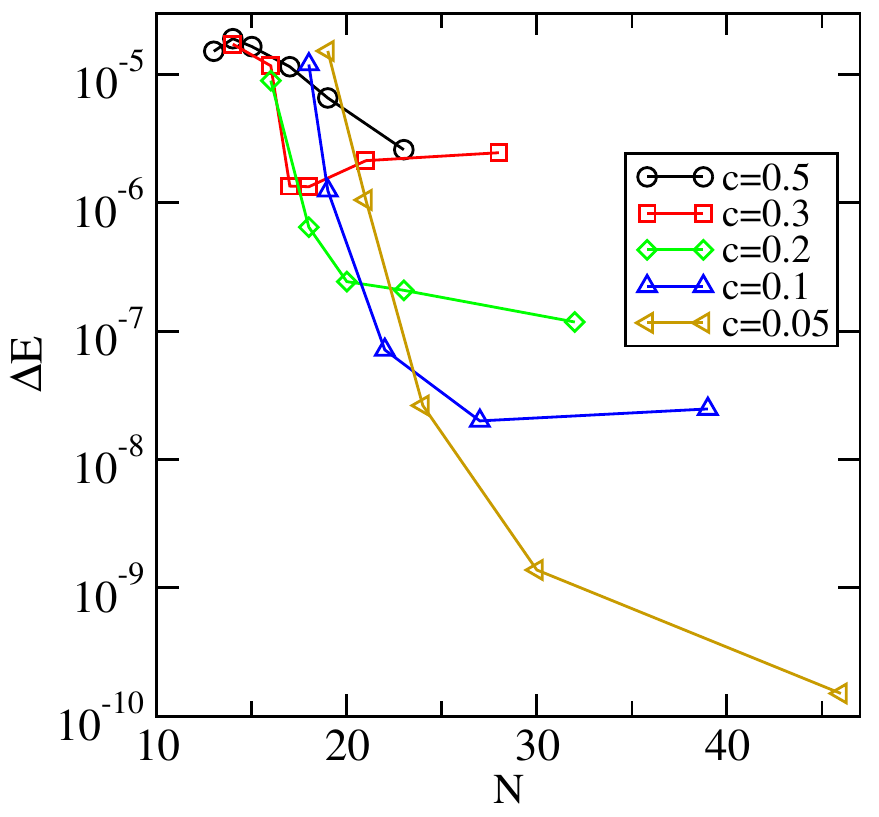}
\caption{Error in the RHF energy of the He atom energy versus number of radial gausslets for $K=6$, with two added x-gaussians. Here $c$ represents the core spacing of the coordinate transformation, and the symbols correspond to different values of $s$, ranging from $0.7$ to $0.05$. Radial functions are omitted if their center is at radius greater than $10$. The actual spacing of functions near $r=0$ is about an order of magnitude smaller than $c$,  obtained by shrinking the functions of Fig. 3 by a factor of $c$. }
\label{fig:errorHe}
\end{figure}

Figure \ref{fig:errorHe} shows the errors in the RHF energy of the He atom given the finite basis and IDA approximated interaction, relative to the precise value -2.8616799956122... of Cinal\cite{Cinal2020} using a pseudospectral method.
We see that microHartree accuracy is obtained with less than 20 functions and about $10^{-9}$ accuracy with about 30.  Compared to a conventional basis where interactions would involve a four index tensor, the radial gausslets with their two-index interactions are much more efficient. This is the ``sweet spot'' in this type of basis construction: only a modest increase in the size of the basis compared to optimal, fully delocalized functions, but with a highly accurate diagonal approximation making calculations much faster. 

\subsection{Introducing spherical harmonics and multipole IDA}
\label{sec:sphericalharm}

To form a more general atomic basis we combine the radial gausslets with spherical harmonics.  For the present
paper we focus on the standard $Y_{\ell m}$ route; later we will also consider localized angular
functions.  A one--electron basis function has the product form
\begin{equation}
\phi_{a\ell m}(\br)
\equiv \frac{\chi_a(r)}{r}\,Y_{\ell m}(\Omega),
\qquad \br \equiv (r,\Omega),
\end{equation}
where $\{\chi_a(r)\}$ are the orthonormal radial gausslets constructed above (including the
$u(0)=0$ boundary condition), and we take real spherical harmonics for simplicity.
We choose an angular cutoff $\ell \le \ell_{\max}$ and include all $m=-\ell,\ldots,\ell$.

Because the radial and angular parts are orthonormal separately, the full basis is orthonormal.
The one--particle Hamiltonian is also conventional: for fixed $\ell$ it is the same radial matrix
for every $m$, i.e.\ it is block--diagonal in $(\ell,m)$.  In particular, we do \emph{not} use IDA
for any one--particle term.  Writing
\begin{equation}
H^{(1)}(\ell) = T^{(0)} + V^{\rm nuc} + V^{\rm cent}(\ell),
\end{equation}
the centrifugal contribution is
\begin{equation}
V^{\rm cent}_{ab}(\ell)
= \frac{\ell(\ell+1)}{2}\int_0^\infty \frac{\chi_a(r)\chi_b(r)}{r^2}\,dr,
\end{equation}
and the full one--electron matrix elements are
\begin{equation}
\big\langle a\ell m \big|\,H^{(1)}\,\big| b\ell' m' \big\rangle
= \delta_{\ell\ell'}\delta_{mm'}\,H^{(1)}_{ab}(\ell).
\end{equation}
All radial integrals are evaluated on the fine quadrature grid.

\paragraph{Coulomb multipoles and what IDA changes.}
The Coulomb interaction is separated into radial and angular factors using the standard multipole expansion
\begin{equation}
\frac{1}{|\br-\br'|}
=
\sum_{L=0}^{\infty}\sum_{M=-L}^{L}
\frac{4\pi}{2L+1}\,
K^{(L)}(r,r')\,
Y_{LM}(\Omega)\,Y_{LM}^*(\Omega'),
\end{equation}
with the radial kernel
\begin{equation}
K^{(L)}(r,r') \equiv \frac{r_<^L}{r_>^{L+1}},
\end{equation}
where $r_{<,>} \equiv \big(\min,\max\big)(r,r')$ and the $L$ superscripts represent powers.
With an angular cutoff $\ell\le \ell_{\max}$, the Gaunt selection rules imply that only
multipoles up to $L\le 2\ell_{\max}$ are ever needed, so we build and store the radial objects
$\{V^{(L)}\}$ only for $L=0,1,\ldots,2\ell_{\max}$.

The only nonstandard step is that we apply IDA to the
\emph{radial} part of each multipole.  Define the integrated radial weights
\begin{equation}
w_a^\chi \equiv \int_0^\infty \chi_a(r)\,dr.
\end{equation}
Then for each $L$ we define a symmetric two--index radial matrix
\begin{equation}
\label{eq:VL_def}
V^{(L)}_{ab}
\equiv
\frac{1}{w_a^\chi\,w_b^\chi}
\int_0^\infty\!\!\int_0^\infty
\chi_a(r)\,K^{(L)}(r,r')\,\chi_b(r')\,dr\,dr'.
\end{equation}
This is the direct generalization of the $L=0$ kernel $1/\max(r,r')$ used in the pure--radial helium test.
Conceptually, IDA collapses the radial \emph{pair density} on each electron from a general product
$\chi_a(r)\chi_c(r)$ to an effectively diagonal form in the radial index, leaving the angular couplings exact.
As a result, the interaction retains the full four--index structure in angular labels, but only a
two--index structure in radial labels.

\paragraph{Fast evaluation of the radial double integral.}
A naive evaluation of Eq.~(\ref{eq:VL_def}) would require a two--dimensional quadrature over $(r,r')$.
Instead we use the standard split--domain identity.
Define the prefix integral
\begin{equation}
J_q^{(L)}(r) \equiv \int_0^{r} q(s)\,s^{L}\,ds,
\end{equation}
where $q$ denotes some function of $s$. Then
\begin{equation}
\label{eq:split_identity}
\begin{aligned}
&\iint_0^\infty f(r)\,g(r')\,K^{(L)}(r,r')\,dr\,dr'
\\
&=\int_0^\infty \frac{f(r)}{r^{L+1}}\,J_g^{(L)}(r)\,dr
+
\int_0^\infty \frac{g(r)}{r^{L+1}}\,J_f^{(L)}(r)\,dr .
\end{aligned}
\end{equation}
so that on a grid the inner integrals become simple cumulative sums (prefix integrals).  This reduces
the cost from $\mathcal O(N_g^2)$ to $\mathcal O(N_g)$ per pair $(a,b)$, and the resulting $V^{(L)}$ is
symmetric by construction. 

\paragraph{Angular factors (Gaunt coefficients).}
It is convenient to combine the angular quantum numbers into a single compound index
$\mu \equiv (\ell_\mu,m_\mu)$, so that a one--electron basis function is labeled by $(a,\mu)$.
The angular dependence of the Coulomb interaction is then expressed in terms of Gaunt coefficients
\begin{equation}
G^{LM}_{\mu\kappa}
\equiv
\int d\Omega\;
Y_{\mu}(\Omega)\,Y_{LM}(\Omega)\,Y_{\kappa}(\Omega),
\end{equation}
(with complex conjugation omitted, since we use real spherical harmonics).
These coefficients satisfy the standard triangle/parity/$m$ selection rules and are highly sparse.
%we treat their construction and storage as standard and summarize implementation notes in
%Appendix~\ref{app:gaunt}.

Putting the radial IDA matrices and the Gaunt factors together, the approximate two--electron matrix
elements become

\begin{equation}
\label{eq:eri_ylm_ida}
\begin{aligned}
&\langle a\mu,\; b\nu \,|\, r_{12}^{-1} \,|\, c\kappa,\; d\lambda\rangle
\\
&\qquad\approx
\delta_{ac}\,\delta_{bd}\,
\sum_{L=0}^{2\ell_{\max}}
\frac{4\pi}{2L+1}\,
V^{(L)}_{ab}\,
\Gamma^{(L)}_{\mu\kappa;\nu\lambda}.
\end{aligned}
\end{equation}
with
\begin{equation}
\Gamma^{(L)}_{\mu\kappa;\nu\lambda}
\equiv
\sum_{M=-L}^{L}
G^{LM}_{\mu\kappa}\,
G^{LM*}_{\nu\lambda}.
\end{equation}
Equation~(\ref{eq:eri_ylm_ida}) is the central structural result for the $Y_{\ell m}$ approach:
the expensive radial dependence enters only through the precomputed two--index matrices $V^{(L)}$,
while the remaining angular algebra is handled by sparse Gaunt couplers.  This is precisely the
sense in which the method preserves the benefits of radial gausslets (compactness and two--index
interaction structure in the large radial space) while retaining the familiar spherical--harmonic
angular representation.

%\Cref{app:gaunt} summarizes implementation notes (per--$M$ slices, symmetry, and numerical hygiene).

\begin{table}[t]
\centering
\caption{Unrestricted Hartree--Fock (UHF) total energies (Hartree) for first-row atoms. The values shown are for \emph{both}  MRCHEM and radial gausslets; for the number of digits shown in each case the results were identical. MRCHEM used a bounding box of size $64$~bohr with requested relative precision of $2\times 10^{-10}$ (near the minimum allowed).
The radial gausslet calculations used $s=0.15$ and $c=s/(2Z)$, with maximum orbital angular momentum up to $\ell_{\max}=8$. For Be, the RHF solution is shown; finding the true broken symmetry UHF solution requires careful treatment\cite{Ivanov,White2023}
}
\label{tab:uhf_mrchem_gausslets_firstrow}
\begin{tabular}{l S[table-format=-3.10]}
\hline
Atom & {\text{$E_{\mathrm{UHF}}$ (Ha)}} \\
\hline
Li  & -7.4327509211 \\
Be(RHF)  & -14.573023168 \\
B   & -24.53315846 \\
C   & -37.69374038 \\
N   & -54.404548303 \\
O   & -74.81898015 \\
F   & -99.41630602 \\
Ne  & -128.547098109 \\
\hline
\end{tabular}
\end{table}

As a test of the combination of radial gausslets and spherical harmonics, in Table I we show the unrestricted Hartree--Fock energies for the first row atoms Li-Ne. The only symmetry imposed was that orbitals were either up or down, and not mixed spin.  To assess the accuracy of the gausslets we compared with MRCHEM\cite{mrchem} results at high accuracy.  To illustrate the uniform convergence of the gausslets we used coordinate mapping parameters $s=0.15$ and $c=s/(2Z)$ for all cases, and functions were kept out to a radius of $R=30$ bohr. Both methods are accurate to at least 10 total digits and are limited by numerical precision; comparisons in cases where higher-precision reference results are available indicate that the accuracies of the two approaches are comparable.  The number of radial basis functions ranged from 50 to 58 as $Z$ was varied.  The HF was performed for successively larger maximum $\ell$ in the basis, ensuring convergence with $\ell$ to all digits. The Ne calculation on an M2 Mac desktop took about 30 seconds, most of it in Hamiltonian construction, especially the radial
integrals entering the nuclear-potential and two-electron terms. The MRCHEM run for Ne on the same machine took about 100 minutes.

\begin{figure}[t]
\includegraphics[width=0.8\columnwidth]{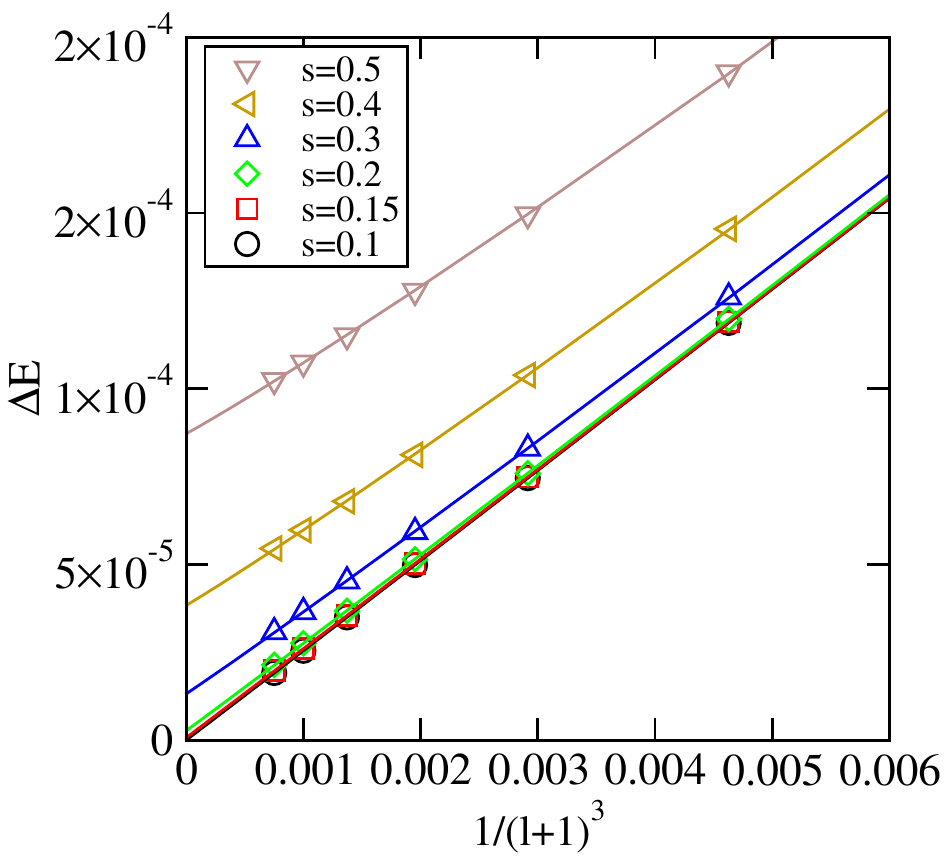}
\caption{Error in the exact diagonalization (full-CI) energy of the He atom versus the maximum angular momentum included for radial gausslets with the indicated distortion parameter $s$, taking $c=s/4$, and including up to $l=10$. The $x$ axis is plotted as $1/(l+1)^3$, the expected leading asymptotic correction. The symbols are the data, and the curves are fits of the form $a + b(l+1)^{-3}+c(l+1)^{-4}$, where $l \le 4$ data was excluded.    }
\label{fig:errorHeFCI}
\end{figure}

In Fig. \ref{fig:errorHeFCI}, we show the error in the fully correlated energy for the He atom as the maximum angular momentum $l$ is varied, for various radial spacing parameters $s$.  Also shown are fits to the data with the expected asymptotic form for a complete radial basis with limited angular momentum\cite{BromleyMitroy07}.  The extrapolated values were then used in a fit of $E(s)$ versus $s$, finding an approximate $s^4$ convergence.  Extrapolating this to $s=0$ gives an error of $1.5\times10^{-7}$ compared to the exact complete basis set limit. This correlated He test shows the radial basis is not just good for HF; its support of the expected angular extrapolation behavior indicates just as good properties for fully correlated calculations.

\section{Discussion and outlook}

We have introduced \emph{radial gausslets}, an orthonormal and localized radial basis that retains the key practical advantage of gausslets: an accurate \emph{diagonal} (two-index) approximation for the Coulomb interaction.  In this construction we have had to adapt gausslets to both the presence of the $r^2$ metric and the half-line domain and boundary conditions at $r=0$.  

We first presented a boundary-gausslet construction which restores completeness and orthogonality near edges.  Turning to radial systems in 3D, we incorporate the metric by working with $u(r)=r R(r)$, but that necessitates the boundary condition $u(0)=0$ to keep the wavefunction finite. Our construction removes the nonzero $r=0$ component from the basis so that no constraint is necessary.  However, this degrades the ability of the basis to satisfy the exact moment conditions that underlie the standard gausslet diagonal approximation.  By adding a small number of additional near-origin functions (the ``$x$-Gaussians''), we nearly restore the moment properties so that an accurate diagonal approximation is available even with a small radial basis. The construction then adds a flexible coordinate mapping which provides systematic control of spatial resolution, concentrating basis functions near the nucleus.

For atomic calculations we combined radial gausslets with spherical harmonics and used the integral-diagonal approximation (IDA) for the radial part of the electron--electron interaction.  This yields an interaction represented by a set of two-index radial matrices $V^{(L)}_{ab}$ rather than a four-index tensor.  The resulting compression is valuable whenever storage or manipulation of four-index Coulomb integrals is a dominant cost.  

The numerical tests demonstrate that radial gausslets provide a compact description of both mean-field and correlated atomic problems.  For example, in a pure radial ($s$-wave) helium test, microhartree accuracy is achieved with fewer than a few dozen radial functions, and substantially higher precision is reachable with only modest additional refinement. No atom-specific optimization was needed; the bases are general-purpose.  For first-row atoms at the Hartree--Fock level, the radial-gausslet results match the accuracy of high precision real-space reference calculations, with a relative accuracy of about 10 digits.  For correlated helium (full-CI in the truncated angular basis), the residual error versus $\ell_{\max}$ is consistent with the expected asymptotic behavior of partial-wave truncation errors.

An attractive feature of our approach is the separation of quadrature from basis representation. Integrals over the radial gausslets use a fine quadrature grid which can capture the singular structure of Coulomb interactions. The basis functions then can have a much coarser grid which suffices for completeness, and which still allows diagonal interactions. 

Several extensions appear promising.  First, the present work used standard spherical harmonics; introducing \emph{localized} angular functions may eventually allow a more fully diagonal interaction representation than the partial form used here. This would be particularly useful for density matrix renormalization group (DMRG) calculations, where the degree of diagonality directly affects the bond dimension of the matrix product operator representation of the Hamiltonian. However, even the partial diagonal form using spherical harmonics should already be useful for DMRG, by substantially reducing the MPO dimension.

Overall, radial gausslets provide an efficient, systematically improvable, and conceptually simple atom-centered basis that bridges grid-like locality and basis-set variational control, while substantially reducing the complexity of Coulomb interaction handling.  We expect these properties to make them useful in a range of atomic and electronic-structure applications where basis size and two-electron integral complexity are key bottlenecks.

\acknowledgments
I thank Sandeep Sharma and Miles Stoudenmire for helpful discussions. This work was supported by the U.S. NSF under Grant
DMR‑2412638. A Julia implementation of the radial gausslet constructions described here is available in the public repository https://github.com/srwhite59/GaussletBases.jl.

\bibliographystyle{apsrev4-2}
\bibliography{ref}

\end{document}